# RVE Analysis in LS-DYNA for High-fidelity Multiscale Material Modeling

Haoyan Wei[*], Dandan Lyu, Wei Hu, C.T. Wu

Computational & Multiscale Mechanics Group, ANSYS Inc.,
Livermore, CA 94551, USA

* Email address: haoyan.wei@ansys.com

## Abstract

In modern engineering designs, advanced materials (e.g., fiber/particle-reinforced polymers, metallic alloys, laminar composites, etc.) are widely used, where microscale heterogeneities such as grains, inclusions, voids, micro-cracks, and interfaces significantly affect the macroscopic constitutive behaviors. Obviously, an accurate description of the multiscale material behaviors is of great importance to the success of material design and structural analysis. The *Representative Volume Element (RVE)* analysis method provides a rigorous means to obtain homogenized macroscopic material properties at the upper length scale from the properties of the material constituents and structures at a lower length scale. Recently, we have developed an RVE module (keyword: *RVE_ANALYSIS_FEM) in the multiphysics simulation software LS-DYNA to enable high-fidelity virtual testing of numerically re-constructed material samples at user-specified characteristic length scales. In this article, a brief introduction to this new feature will be given.

# 1. Basic Theory of RVE Analysis

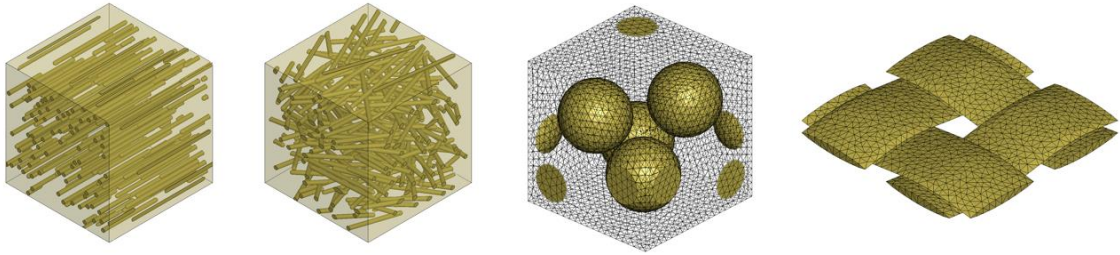

Figure 1. Illustration of the Representative Volume Element (RVE) for composites

RVE is a material volume that statistically represents a sampling of microstructural heterogeneities included at a material point of a continuum body [1], see Figure 1 for illustrations. The size of RVE should be chosen to be large enough so that any volume of an increased size can be considered equally representative. Let us consider an RVE (refer to Figure 2 for a 2D illustration) in its undeformed initial configuration $\Omega$, for which the external boundary can be expressed as $\partial \Omega = \cup_{\alpha=1}^{d} \partial \Omega_\alpha$, where $d = 2$ for 2D RVE models, and $d = 3$ for 3D RVE models. Further, we use $\boldsymbol{X}_\alpha^+ \in \partial \Omega_\alpha^+$ and $\boldsymbol{X}_\alpha^- \in \partial \Omega_\alpha^-$ to denote the microscale material points located on a pair of opposite external surfaces $\partial \Omega_\alpha^+$ and $\partial \Omega_\alpha^-$ of the RVE, respectively.

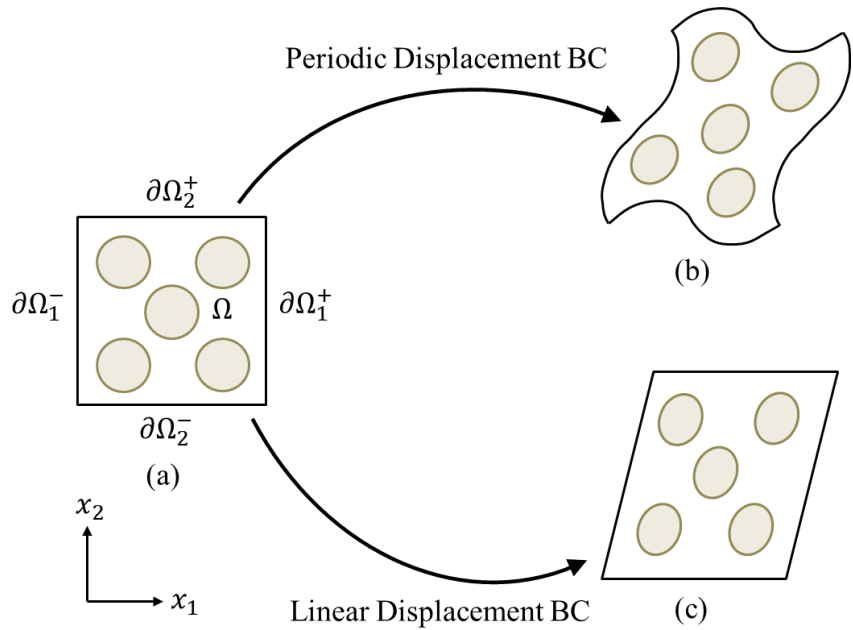

**Figure 2**. Schematic of a 2D RVE and different displacement boundary conditions. (a) The initial configuration of RVE occupying the domain $\Omega$, where $\partial \Omega_\alpha^+$ and $\partial \Omega_\alpha^-$ denote a pair of opposite external boundaries, and $\alpha = 1, 2$ for this 2D model. (b) The current configuration of RVE subjected to periodic displacement boundary conditions. (c) The current configuration of RVE subjected to linear displacement boundary conditions.

In the RVE domain $\Omega$, the microscopic stress $\boldsymbol{\sigma}(X)$ at any point $X \in \Omega$ satisfies the following equilibrium equation:

$$\nabla_X \cdot \boldsymbol{\sigma}(X) = \mathbf{0}, \tag{1}$$

where $\nabla_X$ denotes the gradient operator with respect to the micro-scale. For the microscopic material constituents, constitutive laws can be introduced to define the relationship between the incremental stress $\Delta\boldsymbol{\sigma}(X)$ and strain $\Delta\boldsymbol{\varepsilon}(X)$ as follows:

$$\Delta\boldsymbol{\sigma}(X) = \boldsymbol{C}(X) : \Delta\boldsymbol{\varepsilon}(X), \tag{2}$$

where $\boldsymbol{C}(X)$ denotes the microscopic material stiffness tensor at $X \in \Omega$, and the microscopic strain $\boldsymbol{\varepsilon}(X)$ can be further expressed as a function of the microscopic displacement field $\boldsymbol{w}(X)$ based on the kinematic compatibility condition.

Taking the volume average of the microscopic stress $\boldsymbol{\sigma}$ over the volume $\Omega$, we obtain the homogenized macroscopic stress $\widetilde{\boldsymbol{\sigma}}$

$$\widetilde{\boldsymbol{\sigma}} = \frac{1}{\Omega} \int_\Omega \boldsymbol{\sigma}(X) \, d\Omega, \tag{3}$$

Similarly, the homogenized macroscopic displacement gradient $\widetilde{\boldsymbol{H}}$ is linked to the microscopic displacement gradient $\nabla_X \boldsymbol{w}(X)$ as follows

$$\widetilde{\boldsymbol{H}} = \frac{1}{\Omega} \int_\Omega \nabla_X \boldsymbol{w}(X) \, d\Omega, \tag{4}$$

where $\widetilde{\boldsymbol{H}}$ is defined as

$$\widetilde{\boldsymbol{H}} \equiv \nabla_{\widetilde{X}} \widetilde{\boldsymbol{u}} = \widetilde{\boldsymbol{F}} - \boldsymbol{I}, \tag{5}$$

in which $\nabla_{\widetilde{X}}$ denotes the gradient operator with respect to the macro-scale, $\widetilde{\boldsymbol{u}}$ denotes the macroscopic displacement field, $\widetilde{\boldsymbol{F}}$ denotes the macroscopic deformation gradient, and $\boldsymbol{I}$ denotes an identity tensor.

To describe the macroscopic constitutive behavior, we need to compute the homogenized stress $\widetilde{\boldsymbol{\sigma}}$ and displacement gradient $\widetilde{\boldsymbol{H}}$, which requires the calculation of microscopic stress and displacement fields by solving the equilibrium Eq. (1). To this end, properly defined boundary conditions must be imposed on the RVE boundary. Depending on the relationship between the macroscale deformation $\widetilde{\boldsymbol{H}}$ and the microscale displacement field $\boldsymbol{w}$, two types of displacement boundary conditions are commonly adopted.

The *linear displacement boundary condition (LDBC)* can be specified such that the microscale displacement on the RVE boundary is defined as

$$\boldsymbol{w}_\alpha(X_\alpha) = \widetilde{\boldsymbol{H}} \cdot X_\alpha, \tag{6}$$

where $X_\alpha \in \partial\Omega_\alpha$ denotes an arbitrary microscale material point on the external boundary of RVE.

Alternatively, if the *Periodic Displacement Boundary Condition* (PDBC) is imposed, the microscale displacements $w_\alpha^+$ and $w_\alpha^-$ of the external boundary material points $X_\alpha^+$ and $X_\alpha^-$ are enforced to satisfy the following condition:

$$w_\alpha^+(X_\alpha^+) - w_\alpha^-(X_\alpha^-) = \widetilde{H} \cdot (X_\alpha^+ - X_\alpha^-), \qquad (7)$$

where we have decomposed $\partial\Omega_\alpha$ into a pair of opposite external surfaces $\partial\Omega_\alpha^+$ and $\partial\Omega_\alpha^-$ of the RVE, respectively, i.e., $\partial\Omega_\alpha = \partial\Omega_\alpha^+ \cup \partial\Omega_\alpha^-$, $X_\alpha^+ \in \partial\Omega_\alpha^+$, and $X_\alpha^- \in \partial\Omega_\alpha^-$, as illustrated in Figure 2.

***Remark 2.1.*** It is noteworthy to mention that RVEs with LDBC usually appear to be stiffer than RVEs with PDBC. When the size of RVE is large enough, however, the influence of different boundary conditions on the homogenized material properties becomes negligibly small.

***Remark 2.2.*** Traditionally, enforcing PDBC requires the finite element mesh to be PDBC-matching, which means that the nodal distributions on the RVE's opposite sides must match well with each other. For instance, let us consider two opposite surfaces (e.g., $\partial\Omega_1^+$ and $\partial\Omega_1^-$) that are both perpendicular to the $x_1$ axis, if we draw a straight line that starts from any FEM node on surface $\partial\Omega_1^+$ and is parallel to the $x_1$ axis, then a node must be present at the intersection point of this line with surface $\partial\Omega_1^-$. The same condition applies if we draw a line from any node on $\partial\Omega_1^-$ to the surface $\partial\Omega_1^+$.

***Remark 2.3.*** In our implementation in LS-DYNA [2], both PDBC-matching and non-matching meshes can be used to impose the periodic displacement boundary condition. For some cases, the non-matching mesh is attractive because the creation of PDBC-matching meshes could be tedious if very complex material micro-structures exist. Nevertheless, a PDBC-matching mesh is preferred for most cases, as it allows for the use of very efficient and accurate homogenization algorithms.

***Remark 2.4.*** Enforcing PDBC/LDBC to an RVE model used to be a non-trivial task for most commercial FEA software users, as manually specifying multiple constraint conditions is quite tedious and error-prone. To offer a more user-friendly experience, we have designed the keyword *RVE_ANALYSIS_FEM in LS-DYNA in such a way that the users only need to choose PDBC or LDBC, and the software will automatically set up all the constraint conditions for the RVE.

## 2. Application based on RVE Analysis

RVE analysis becomes increasingly useful for advanced material design and structural analysis. Although there are many different application areas, in this section we will only briefly discuss two RVE-based applications.

### 2.1. Parameter Identification for Constitutive Laws

Constitutive models often involve many parameters, especially for nonlinear anisotropic materials. Calibration of these model parameters requires many material data. However, measuring these data from a series of physical experiments is quite time-consuming and expensive. These physical experiments, however, can be easily replaced by more cost-saving numerical RVE analysis. By performing homogenization of properly designed material samples, RVE models can accurately predict the macro-scale material behaviors, which can then be adopted to identify the material parameters of the assumed constitutive model [3].

### 2.2. Multiscale Structural Simulation

To perform multiscale structural simulation, separate finite element (FE) models are needed to represent structures at different length scales. Taking the multiscale car crash simulation as an example, we can construct a macro-scale FE model to represent a car body made of composite materials, whereas a micro-scale FE model (i.e., RVE model) is built to represent the material microstructures. At each integration point of the macro-scale FE model, there is an associated FE-based RVE model, so this multiscale structural simulation method is often referred to as the 'FE$^2$ method'. During a concurrent multiscale simulation, the stress/deformation state of the macro-scale FE model serves as the boundary condition for the RVE model, and the micro-scale RVE simulation yields the macroscopic material behaviors for the macro-scale FE model. This two-way macro-micro coupling occurs at each time step, and thus there is no need to specify any constitutive laws at the macroscopic level.

This concurrent simulation offers high-fidelity multiscale structural simulation, but the resulting computational costs can be very high for large-scale engineering problems. Therefore, a reduced-order multiscale model that achieves both high accuracy and efficiency is desired. To this end, we are currently developing a mechanistic machine learning-based multiscale simulation approach for composite structures. In this approach, we employ material data from high-fidelity RVE simulations to perform offline training of machine learning models named Deep Material Network (DMN) [4]. Replacing high-fidelity RVE with DMN in a concurrent multiscale structural simulation can offer a computational speed orders-of-magnitude faster than the conventional FE$^2$ method. Currently, this DMN-based multiscale simulation method is under development in LS-DYNA for injection-molded short-fiber-reinforced composite structures [5, 6].

## 3. Workflow for RVE Analysis in LS-DYNA

### 3.1 RVE Analysis Procedure

To prepare the input deck for RVE analysis in LS-DYNA, the following steps and associated keywords should be considered:

**(1) RVE Construction** *(\*SECTION_xxx, \*PART, \*ELEMENT_xxx, \*NODE):*

> a. define the microstructure geometry
> b. generate finite element mesh

**(2) Microscopic Material Model** *(\*MAT_xxx):*

> a. define the constitutive model for each material constituent (*Eq. (2)*)
> b. provide the corresponding material constants

**(3) Boundary Condition** *(\*RVE_ANALYSIS_FEM, \*DEFINE_CURVE, \*CONTROL_TERMINATION):*

> a. choose LDBC (*Eq. (6)*) or PDBC (*Eq. (7)*)
> b. define the loading history with displacement gradient $\widetilde{H}$ (*Eq. (5)*)

**(4) Output Control** *(\*DATABASE_RVE, \*DATABASE_BINARY_D3PLOT):*

> a. define the output parameters for the d3plot file
> b. define the output frequency for the rveout file

**(5) Implicit Solver** *(\*CONTROL_IMPLICIT_GENERAL, \*CONTROL_IMPLICIT_AUTO, \*CONTROL_IMPLICIT_SOLUTION, \*CONTROL_IMPLICIT_SOLVER, etc.):*

> a. choose a direct or iterative solver
> b. define the implicit solver parameters

After the input deck is prepared, the RVE analysis job can be submitted in the same way as other standard SMP/MPP LS-DYNA simulations.

When the simulation is finished, 'd3plot' files will be generated, which contain the microscopic stress/deformation fields of the RVE, and these data can be visualized directly in the LS-PrePost software. Meanwhile, an 'rveout' file containing time histories of homogenized macroscopic stress/deformation fields is also created, and this file can be opened by any text editors.

*Remark 4.1.* While *RVE_ANALYSIS_FEM and *DATABASE_RVE are the most important keywords for the RVE model definition, a successful RVE analysis also requires the proper definition of other keywords (including keywords that are not listed in this article) to ensure the efficiency and accuracy of FEM simulation.

***Remark 4.2.*** From the numerical point of view, a good mesh quality is required to obtain accurate numerical results. As in any finite element analysis, the finite element size should be small enough to ensure that numerically converged results are obtained.

***Remark 4.3.*** From the micromechanics point of view, the size of the RVE should be large enough to ensure that accurate homogenization results are obtained.

### 3.2 RVE Example

As an example, let us consider a 3D RVE model for spherical particle-reinforced composites. Firstly, we follow the procedure described in Section 4.1 to prepare the input deck for RVE analysis.

In Step (1), we construct the RVE model based on the particle geometry and volume fraction. Figure 3 shows the PDBC-matching finite element mesh consisted of particle phase and matrix as two parts (*PART) perfectly bonded together. The mesh information (*ELEMENT_SOLID, *NODE) is stored in the file 'mesh3d_particle_composite.k'.

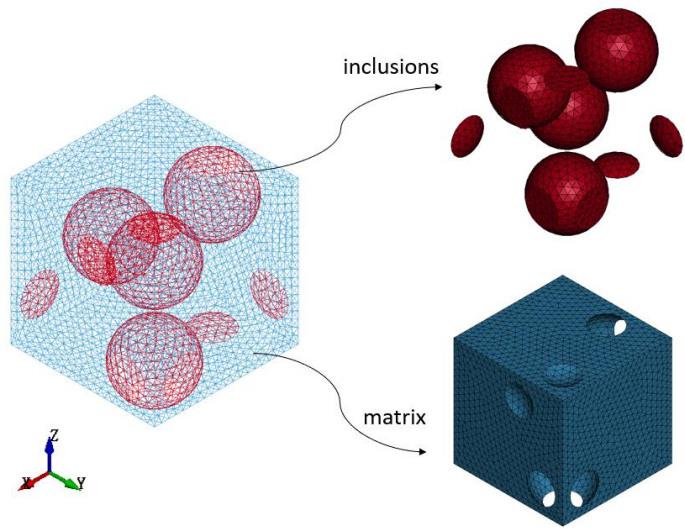

**Figure 3**. The RVE model for spherical particle-reinforced composites, for which the 10-node tetrahedron solid finite elements are employed

In Step (2), we choose a nonlinear hyperelastic constitutive model with the keyword *MAT_MOONEY-RIVLIN_RUBBER and assign different material constants for the particle phase and the matrix phase, respectively, while different constitutive models can be assigned to each microscopic constituent in general applications.

In Step (3), we define the keyword *RVE_ANALYSIS_FEM as follows:

```
*RVE_ANALYSIS_FEM
mesh3d_particle_composite.k
$---+----1----+----2----+----3----+----4----+----5----+----6-
$    inpt      oupt      lcid      idof        bc      imatch
       0         1         1         3          0         1
$    H11       H22       H33       H12        H23       H13
      0.5
```

where the RVE mesh file name 'mesh3d_particle_composite.k' is given; inpt=0 is used to indicate that the RVE boundary conditions are imposed automatically by LS-DYNA; oupt=1 means that the RVE homogenization results will be written to the 'rveout' file; idof=3 is used for this 3D model; bc=0 means that PDBC will be imposed; imatch=1 is defined because a PDBC-matching mesh is used; and for the macroscopic displacement gradient $\tilde{H}$, its 11 component is defined to be 0.5, while all the other components of $\tilde{H}$ are set free by simply leaving the entries in the input card blank. In addition, lcid=1 in *RVE_ANALYSIS_FEM indicates that the RVE boundary condition associated with the imposed $\tilde{H}$ will be applied incrementally based on the load curve with lcid=1 defined in *DEFINE_CURVE:

```
*DEFINE_CURVE
1
0.0,0.0
1.0,1.0
```

In Step (4), we define the output frequency for d3plot and rveout files as follows:

```
*DATABASE_BINARY_D3PLOT
$  dt/cycl
       0.1

*DATABASE_RVE
$       dt
       0.1
```

In Step (5), we choose to use the direct implicit solver for this example.

```
*CONTROL_TERMINATION
$   endtim    endcyc     dtmin
       1.0         0       0.0
*CONTROL_ACCURACY
$      osu       inn
         1         1
*CONTROL_IMPLICIT_GENERAL
         1, 0.1,,,1
*CONTROL_IMPLICIT_SOLUTION
  , , ,1e-6, 1e-8, , ,1e-30
*CONTROL_IMPLICIT_SOLVER
                 6,0
```

An iterative implicit solver can be adopted to achieve better efficiency than the direct solver, but sometimes it may require some trial-and-error to find the optimal solver parameters.

Except for the mesh information (*ELEMENT_SOLID, *NODE) stored in 'mesh3d_particle_composite.k', we will store all the other keywords in a separate input file named 'input_rve3d.k', and then we can use the following command to run MPP LS-DYNA simulation:

```
mpirun -np 10 lsdyna_mpp_dp i= input_rve3d.k
```

where we have assumed that the executable file is named 'lsdyna_mpp_dp'.

After the finite element simulation is finished, we can open the d3plot file in LS-PrePost to check the microscopic stress and deformation, as shown in Figure 4.

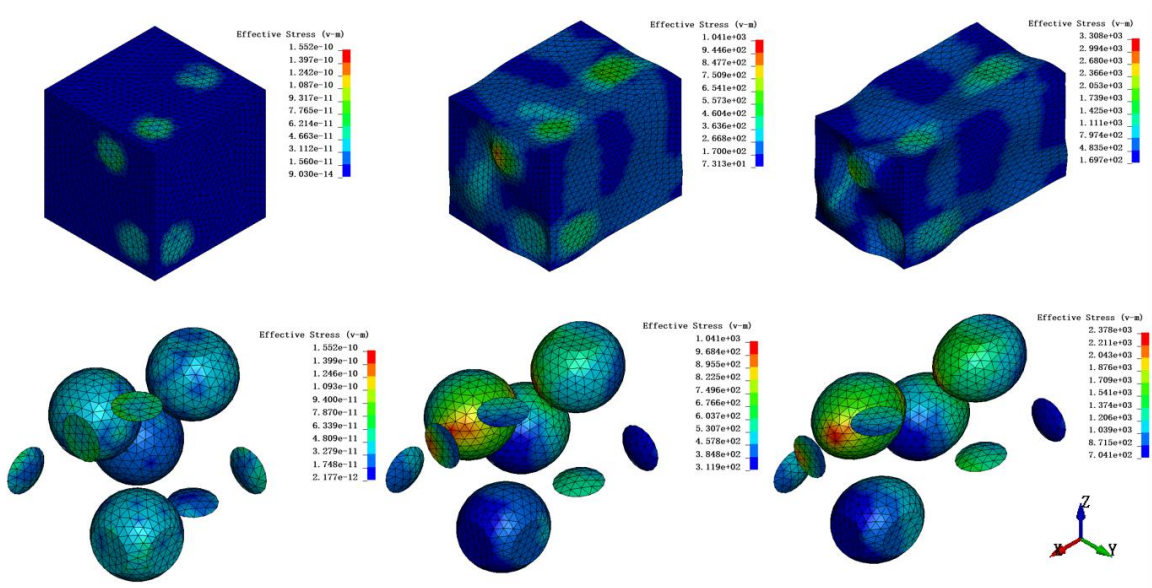

**Figure 4**. RVE simulation results for particle-reinforced composites. The distribution of von Mises stress in the composites and the particle phase are plotted at different loading steps.

In addition, the homogenized macroscopic stress/strain results are available in the rveout file. For instance, the deformation gradient, the Green strain, and the 1st Piola-Kirchhoff (PK1) stress at the last output step is given in rveout as follows:

```
deformation gradient F, Green strain E, PK1 stress P

F11=  0.1500000000E+01   F22=  0.8164978140E+00   F33=  0.8219332698E+00   F12= -0.9944214154E-03   F23= -0.2224868009E-02   F31= -0.1304440114E-02
E11=  0.6250013452E+00   E22= -0.1666626904E+00   E33= -0.1622095242E+00   E12= -0.1150336414E-02   E23= -0.1821997870E-02   E31= -0.1513305221E-02
P11=  0.4641744224E+03   P22=  0.0000000000E+00   P33=  0.0000000000E+00   P12=  0.0000000000E+00   P23=  0.0000000000E+00   P31=  0.0000000000E+00
```

***Remark 4.4.*** The complete input deck for this example can be downloaded from the website: https://www.lstc-cmmg.org/ex-rve

## 4. Conclusion

An RVE module has been developed in the multiphysics simulation software LS-DYNA. Given the material microstructural information (geometry and constitutive properties for base materials), this new feature enables LS-DYNA to perform high-fidelity RVE analysis in an efficiency manner, as it includes the following functions:

(1) Creation of periodic displacement boundary conditions (or linear displacement boundary conditions) automatically;

(2) Nonlinear implicit finite element analysis of RVE under any loading conditions;

(3) Calculation of the detailed distribution and evolution of microscopic stress/strain fields in the RVE;

(4) Homogenization of the RVE's microscopic material responses to yield the macroscopic effective material responses.

In summary, this function allows virtual testing of numerically re-constructed material samples, which is of great importance to design and analysis of advanced materials such as fiber-reinforced composites, particulate composites, laminar composites, polycrystalline aggregates, single-phase or multiphase porous media, etc., as well as structures made of such composite materials.

## Acknowledgments

The authors would like to acknowledge Dr. Zeliang Liu for the helpful discussions and contributions on the LS-DYNA RVE analysis when he was working at Livermore Software Technology.